\documentclass[showkeys,showpacs]{revtex4}
\usepackage{amsmath}
\usepackage{amssymb}
\usepackage{graphicx}
\usepackage{color}

\begin{document}

\title{
Qualitative analysis of the eigenvalue problem for two coupled Ginzburg-Landau equations
}

\author{Vladimir Dzhunushaliev,$^{1,2}$
\footnote{Email: vdzhunus@krsu.edu.kg}
Vladimir Folomeev$^{2}$
\footnote{Email: vfolomeev@mail.ru}
and
Ratbay Myrzakulov$^{3}$
\footnote{Email: cnlpmyra1954@yahoo.com, cnlpmyra@mail.ru}
}

\affiliation{
$^1$
Institute for Basic Research,
Eurasian National University,
Astana, 010008, Kazakhstan; \\ 
$^2$Institute of Physicotechnical Problems and Material Science of the NAS
of the
Kyrgyz Republic, 265 a, Chui Street, Bishkek, 720071,  Kyrgyz Republic \\
$^3$ Department of General and Theoretical Physics,
Eurasian National University, Astana, 010008, Kazakhstan
}

\begin{abstract}
Eigenvalue problem for two coupled Ginzburg-Landau equations is numerically investigated. The fixed points of corresponding equations system are found. The classification of these points is made. The phase portraits of corresponding ordinary differential equations and the dependence of some parameters of the equations system and the total energy on the initial values are given.
\end{abstract}

\maketitle

\section{Introduction}

Systems with coupled scalar fields hold their certain interest in physical applications.
In particular, such systems are used in  constructing particle models
 and their interactions within the framework of quantum field theory \cite{rajaraman}.
In different aspects such systems have been already repeatedly considered, see for example Refs.
\cite{Bazeia:1995en,Bazeia:1997zp,BezerradeMello:2003ei,Bazeia:2004dh,Vernov:2006dm,Cordero:2007kc,Aref'eva:2009xr}.
 Here we present a qualitative study of one of such systems having the potential energy in the form:
\begin{equation}
\label{poten_sf}
    V(\varphi,\psi)=\frac{\Lambda_1}{4}(\varphi^2-m_1^2)^2+
    \frac{\Lambda_2}{4}(\psi^2-m_2^2)^2+\frac{\Lambda_3}{2}\varphi^2 \psi^2+\text{const},
\end{equation}
where $\varphi,\psi$ are two real scalar fields (usual or phantom/ghost ones),
$\Lambda_1, \Lambda_2, \Lambda_3, m_1, m_2$ are some constants.
This potential had been used earlier by us in treatments of models of cosmological and astrophysical objects in general relativity.
These researches showed that:
(a) for the four-dimensional case there exist regular
spherically and cylindrically
symmetric solutions
 \cite{Dzhunushaliev:2006sy,Dzhunushaliev:2006kd,Dzhunushaliev:2007cs}, and also
cosmological solutions \cite{Dzhunushaliev:2006xh,Folomeev:2007uw}
both for usual and phantom scalar fields;
(b) for the higher dimensional cases
there exist the thick brane  solutions
\cite{Dzhunushaliev:2006vv,Dzhunushaliev:2008zz,Dzhunushaliev:2007rv,Dzhunushaliev:2008hq} supported by usual and phantom scalar fields.

From the physical point of view,
these solutions exist because of the special form of
the interaction potential \eqref{poten_sf} having two local and two global minima.
It means that there are two different vacua.
At the infinity, as the radial coordinate $x\to \pm \infty$, these scalar fields
 are located in that vacuum in which their are in the local minimum. The existence of regular solutions with finite energy is only possible
for certain (eigen) values of parameters of a problem. Such eigenvalue problems were solved by us
using  the shooting method (see for details of the shooting procedure in Ref.~\cite{Dzhunushaliev:2007cs}).

Note that the type \eqref{poten_sf} potential has been also used in the paper \cite{Dzhunushaliev:2007xf} for modeling superconductivity
using two coupled  Ginzburg-Landau equations. When the interaction between the fields in \eqref{poten_sf} is excluded, i.e. at
 $\Lambda_3=0$, one has two uncoupled type Ginzburg-Landau equations. On the other hand with an account taken of the interaction
 there are new effects, in particular, the presence of regular solutions, not present in the case of one Ginzburg-Landau equation.

Use of spherical and cylindrical symmetries in the above papers leads to obtaining the sets of coupled non-autonomous
ordinary differential equations whose qualitative study is quite complicated. Here we consider a simplified problem when
the system with the potential  \eqref{poten_sf} is examined in cartesian coordinates. This allows to get an autonomous
system of two second order ordinary differential equations for which it is possible to perform the qualitative  analysis and estimate
the general behavior of the system.

\section{Qualitative analysis}

Let us consider the physical system with the potential
 \eqref{poten_sf} whose Lagrangian can be presented in the form
\begin{equation}
\label{lagran_sf}
L=\frac{\epsilon_1}{2}\partial_{\mu}\varphi\partial^{\mu}\varphi +\frac{\epsilon_2}{2}\partial_{\mu}\psi\partial^{\mu}\psi-V(\varphi,\psi),
\end{equation}
where $\epsilon_1, \epsilon_2=\pm 1$, and the plus sign refers to usual scalar fields, and the minus sign -- to phantom/ghost ones.
The energy-momentum tensor of the system is:
\begin{equation}
\label{emt_two_fields}
T_i^k=\epsilon_1\partial_{i}\varphi\partial^{k}\varphi +\epsilon_2\partial_{i}\psi\partial^{k}\psi
-\delta_i^k L.
\end{equation}
The corresponding field equations in cartesian coordinates can be written as:
\begin{eqnarray}
\label{sf1}
\varphi^{\prime\prime}=\frac{1}{\epsilon_1}\frac{\partial V}{\partial \varphi},
\\
\label{sf2}
\psi^{\prime\prime}=\frac{1}{\epsilon_2}\frac{\partial V}{\partial \psi}.
\end{eqnarray}
Choosing the initial value of the scalar field
 $\varphi(r=0)=\varphi_0$ and introducing the dimensionless variables
 $\phi=\varphi/\varphi_0, \chi=\psi/\varphi_0, x=\sqrt{\Lambda_3}\varphi_0 r, \mu_1=m_1/\varphi_0$,
$\mu_2=m_2/\varphi_0$,
and also
$\lambda_1=\Lambda_1/\Lambda_3, \lambda_2=\Lambda_2/\Lambda_3$,
let us rewrite the system in the form
(taking into account the expression for the potential from \eqref{poten_sf}):
\begin{eqnarray}
\label{sf1_fix}
&&\phi^\prime=z,
\\
\label{sf2_fix}
&&\chi^\prime=v,
\\
\label{sf3_fix}
&&z^\prime=\frac{1}{\epsilon_1}\phi\left[\chi^2+\lambda_1\left(\phi^2-\mu_1^2\right)\right],
\\
\label{sf4_fix}
&&v^\prime=\frac{1}{\epsilon_2}\chi\left[\phi^2+\lambda_2\left(\chi^2-\mu_2^2\right)\right].
\end{eqnarray}
The fixed points of this system are:
\begin{eqnarray}
\label{fix1}
&&A=\left\{z\to 0, v \to 0, \chi \to 0, \phi \to \mu_1\right\},
\\
\label{fix2}
&&B=\left\{z\to 0, v \to 0, \chi \to 0, \phi \to -\mu_1\right\},
\\
\label{fix3}
&&C=\left\{z\to 0, v \to 0, \phi \to 0, \chi \to \mu_2\right\},
\\
\label{fix4}
&&D=\left\{z\to 0, v \to 0, \phi \to 0, \chi \to -\mu_2\right\},
\\
\label{fix5}
&&E=\left\{z\to 0, v \to 0, \phi \to 0, \chi \to 0\right\},
\\
\label{fix6}
&&F=\left\{z\to 0, v \to 0, \phi \to \pm \sqrt{\frac{\lambda_2 \mu_2^2-\lambda_1\lambda_2 \mu_1^2}
           {1-\lambda_1\lambda_2}},
           \chi \to \pm
     \sqrt{\frac{\lambda_1 \mu_1^2-\lambda_1\lambda_2 \mu_2^2}
           {1-\lambda_1\lambda_2}}\right\}.
\end{eqnarray}
The points $A, B$ refer to local minima, the points
 $C, D$ are the global minima,
 $E$ is the local maximum, and the points
 $F$ refer to saddle points. Let us designate the values of the potential
 \eqref{poten_sf} at these points as
 $V_i$ where the index $i$ corresponds to letters from
 $A$ to $F$, and
$V_A=V_B$ and $V_C=V_D$.

Obviously that
\begin{equation}
\label{cond_max}
{\rm max} (V_A,V_C)\leq V_E.
\end{equation}
It is also possible to find the following relations:
\begin{eqnarray}
 &&  V_F-V_A
   =\frac{\lambda_1\big(- \mu_1^2+\lambda_2 \mu_2^2 \big)^2}
         {4(1-\lambda_1\lambda_2)},
\nonumber\\
&&V_F-V_C
   =\frac{\lambda_2\big(- \mu_2^2+\lambda_1 \mu_1^2 \big)^2}
         {4(1-\lambda_1\lambda_2)},
\nonumber\\
&&V_F-V_E=\frac{\lambda_1 \lambda_2
   \big[\mu_1^2 (\lambda_1 \mu_1^2- \mu_2^2)
       +\mu_2^2 (\lambda_2 \mu_2^2- \mu_1^2)\big]}
         {4(1-\lambda_1\lambda_2)}.\nonumber
\end{eqnarray}
Dividing the third expression by the first one and the second one and taking into account the condition
\eqref{cond_max}, one can show that the conditions guaranteeing the existence  of the local and global minima are:
(i) for the local minimum:
 $\lambda_1>0, \mu_1^2>\lambda_2 \mu_2^2$;
 (ii) for the global minimum:
 $\lambda_2>0, \mu_2^2>\lambda_1 \mu_1^2$.

Next, to determine the type of the fixed points it is necessary to make a characteristic matrix of the system
 \eqref{sf1_fix}-\eqref{sf4_fix}. Using this matrix, one can write out the characteristic forth order algebraic equation.
Then the values of roots of this equation $k_j$ determine the type of the fixed points. In our case we have the following roots:

(i) at the points  $A, B$ (the local minimum):
$$
k_1=\frac{2}{\epsilon_1} \lambda_1 \mu_1^2, \quad k_2=\frac{1}{\epsilon_2}\left( \mu_1^2-\lambda_2 \mu_2^2\right),
\quad k_{3,4}=1.
$$

(ii) at the points $C, D$ (the global minimum):
$$
k_1=\frac{1}{\epsilon_1} \left( \mu_2^2-\lambda_1 \mu_1^2\right), \quad
k_2=\frac{2}{\epsilon_2}\lambda_2 \mu_2^2,
\quad k_{3,4}=1.
$$

(iii) at the point $E$ (the local maximum):
$$
k_1=-\frac{1}{\epsilon_1} \lambda_1 \mu_1^2, \quad
k_2=-\frac{1}{\epsilon_2}\lambda_2 \mu_2^2,
\quad k_{3,4}=1.
$$

(iv) at the points $F$ (the saddle points):
\begin{eqnarray}
&&k_{1,2}=-\frac{1}{\epsilon_1 \epsilon_2 (\lambda_1\lambda_2-1)}
\Big\{-\epsilon_2 \mu_1^2\lambda_1^2\lambda_2-\epsilon_1 \mu_2^2\lambda_1\lambda_2^2+\epsilon_1 \mu_1^2\lambda_1\lambda_2+
\epsilon_2 \mu_2^2 \lambda_1\lambda_2 \nonumber \\
&&\pm\sqrt{\lambda_1\lambda_2\left\{4\epsilon_1\epsilon_2(\mu_2^2\lambda_2-\mu_1^2)(\mu_2^2-\lambda_1 \mu_1^2)
(\lambda_1\lambda_2-1)+\lambda_1\lambda_2\left[\mu_1^2(\epsilon_2\lambda_1-\epsilon_1)+
\mu_2^2(\epsilon_1\lambda_2-\epsilon_2)\right]^2\right\}}\,\Big\}, \nonumber \\
&&k_{3,4}=1.\nonumber
\end{eqnarray}

Taking into account the above mentioned conditions of existence of the local and global minima, one can  make the classification of
the fixed points presented in table \ref{tabl1}.

\begin{table}[htbp]
 \caption{\small The classification of the fixed points of the system
 \eqref{sf1_fix}-\eqref{sf4_fix} for different values of the parameters
 $\epsilon_1$ and $\epsilon_2$.}
\vspace{-.7cm}
\begin{center}
\begin{tabular}{|p{2.5cm}|p{2.5cm}|p{2.5cm}|p{2.5cm}|p{3.5cm}|}
\multicolumn{5}{c}{} \\[10pt]\hline
 & Points A, B & Points C, D & Point E &   Point F \\
\hline
$\epsilon_1=1, \epsilon_2=1$&Unstable node&Unstable node& Saddle&Depends on the values of the parameters $\lambda_1,\lambda_2$\\
\hline
$\epsilon_1=1, \epsilon_2=-1$&Saddle&Saddle& Saddle&Depends on the values of the parameters $\lambda_1,\lambda_2$\\
\hline
$\epsilon_1=-1, \epsilon_2=1$&Saddle&Saddle& Saddle&Depends on the values of the parameters $\lambda_1,\lambda_2$\\
\hline
$\epsilon_1=-1, \epsilon_2=-1$&Saddle&Saddle& Unstable node&Depends on the values of the parameters $\lambda_1,\lambda_2$\\
\hline
\end{tabular}
\end{center}
\label{tabl1}
\end{table}

\section{Numerical analysis}

From the point of view of obtaining a set of solutions (but not only of two integral curves as in a case of  saddle fixed points)
  type ``unstable node'' fixed points  are more interesting.
In the model under consideration, they are situated:
 (i) at the points of the local
($A, B$) and global ($C, D$) minima at positive
 $\epsilon_1, \epsilon_2$, i.e. for usual scalar fields;
 (ii) at the point of the local maximum
 $E$ at negative
 $\epsilon_1, \epsilon_2$, i.e. for the case of phantom/ghost fields. Note that in the latter case negative
$\epsilon_1, \epsilon_2$ effectively correspond  to the system with usual fields but with a reversed sign of the potential
\eqref{poten_sf}. In this case the point of the local maximum
$E$ becomes the point of the local minimum, and solutions asymptotically tend to that point.

\begin{table}[htbp]
 \caption{\small The initial values of
 $\chi_0$ and the corresponding values of the parameters
 $\mu_1, \mu_2$ for the system
\eqref{sf1_fix}-\eqref{sf4_fix}.}
\vspace{-.7cm}
\begin{center}
\begin{tabular}{|p{1cm}|p{2.5cm}|p{2.5cm}|p{2.5cm}|p{2.5cm}|p{2.5cm}|}
\multicolumn{6}{c}{} \\[10pt]\hline
 $\#$ &$\phi_0$ & $\chi_0$ & $\mu_1$ & $\mu_2$& $M$ \\
\hline
1&1.0&0.3&1.25104535& 1.1056305&0.0441854\\
\hline
2&1.0&$\sqrt{0.2}$&1.4544857& 1.1878968&0.148259 \\
\hline
3&1.0&$\sqrt{0.4}$&1.736266& 1.30665&0.41906 \\
\hline
4&1.0&$\sqrt{0.6}$&1.9628773& 1.40650056&0.76536 \\
\hline
5&1.0&$\sqrt{0.8}$&2.158048& 1.495301394&1.17074 \\
\hline
6&1.0&1.0&2.33213652& 1.5764432135&1.62578\\
\hline
7&1.0&$\sqrt{1.2}$&2.4908109& 1.6518053896&2.1246\\
\hline
8&1.0&$\sqrt{1.4}$&2.63757479& 1.7225756427&2.66294\\
\hline
\end{tabular}
\end{center}
\label{tabl2}
\end{table}

\begin{figure}[t]
\begin{minipage}[t]{.49\linewidth}
  \begin{center}
  \includegraphics[height=5.5cm]{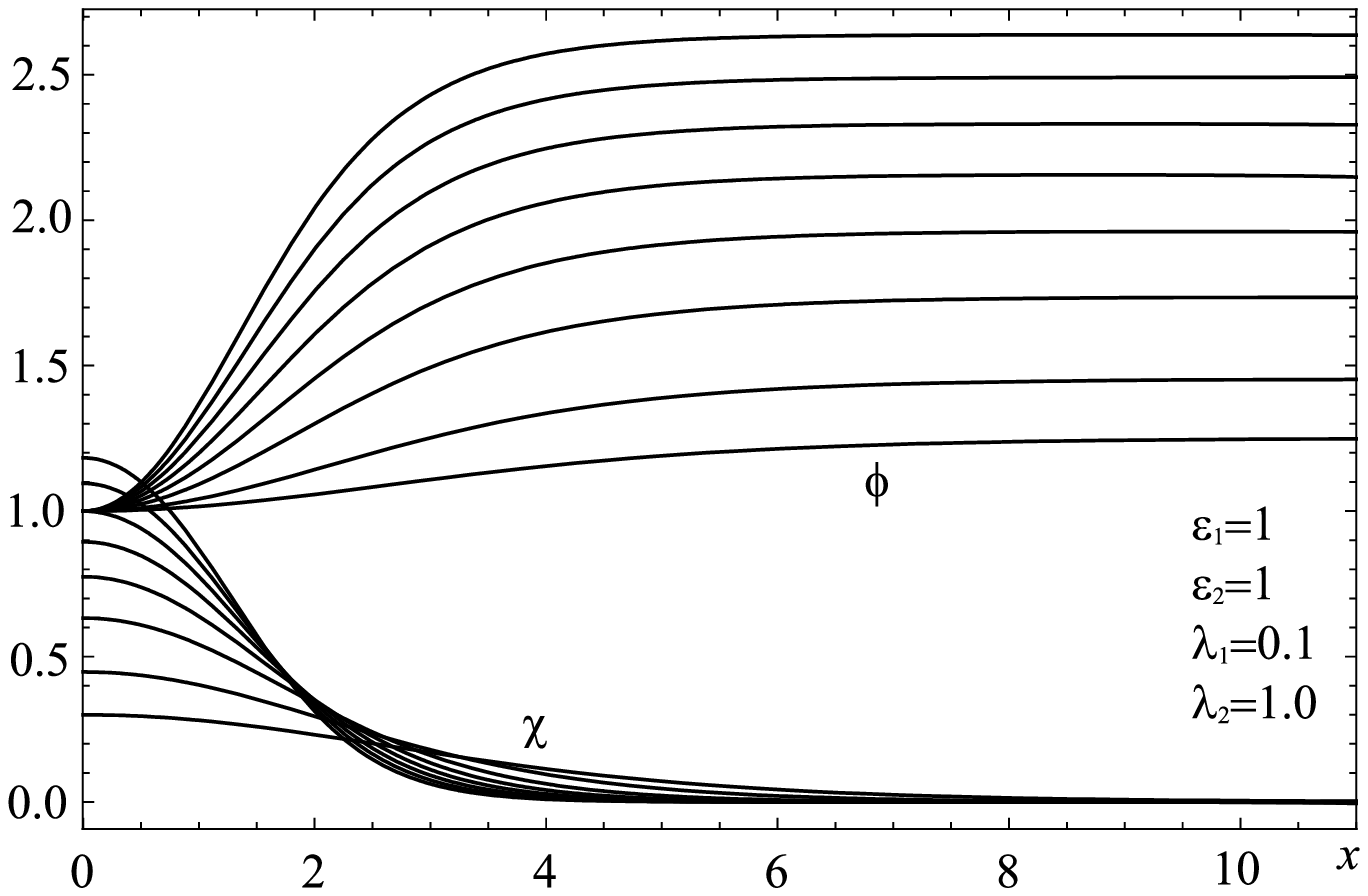}
\vspace{-.2cm}
\caption{\small The scalar fields
 $\phi, \chi$ from the system
  \eqref{sf1_fix}-\eqref{sf4_fix} for the different initial values of
 $\chi_0$ taken from table \ref{tabl2}. Asymptotically, as
 $x\to \pm \infty$, the scalar field
 $\chi \to 0$, and $\phi$ goes to values of
 $\mu_1$ from  table \ref{tabl2} corresponding to the local minimum
$A$ from \eqref{fix1}.
}
    \label{scal_fields}
  \end{center}
\end{minipage}\hfill
\begin{minipage}[t]{.49\linewidth}
  \begin{center}
\includegraphics[height=5.5cm]{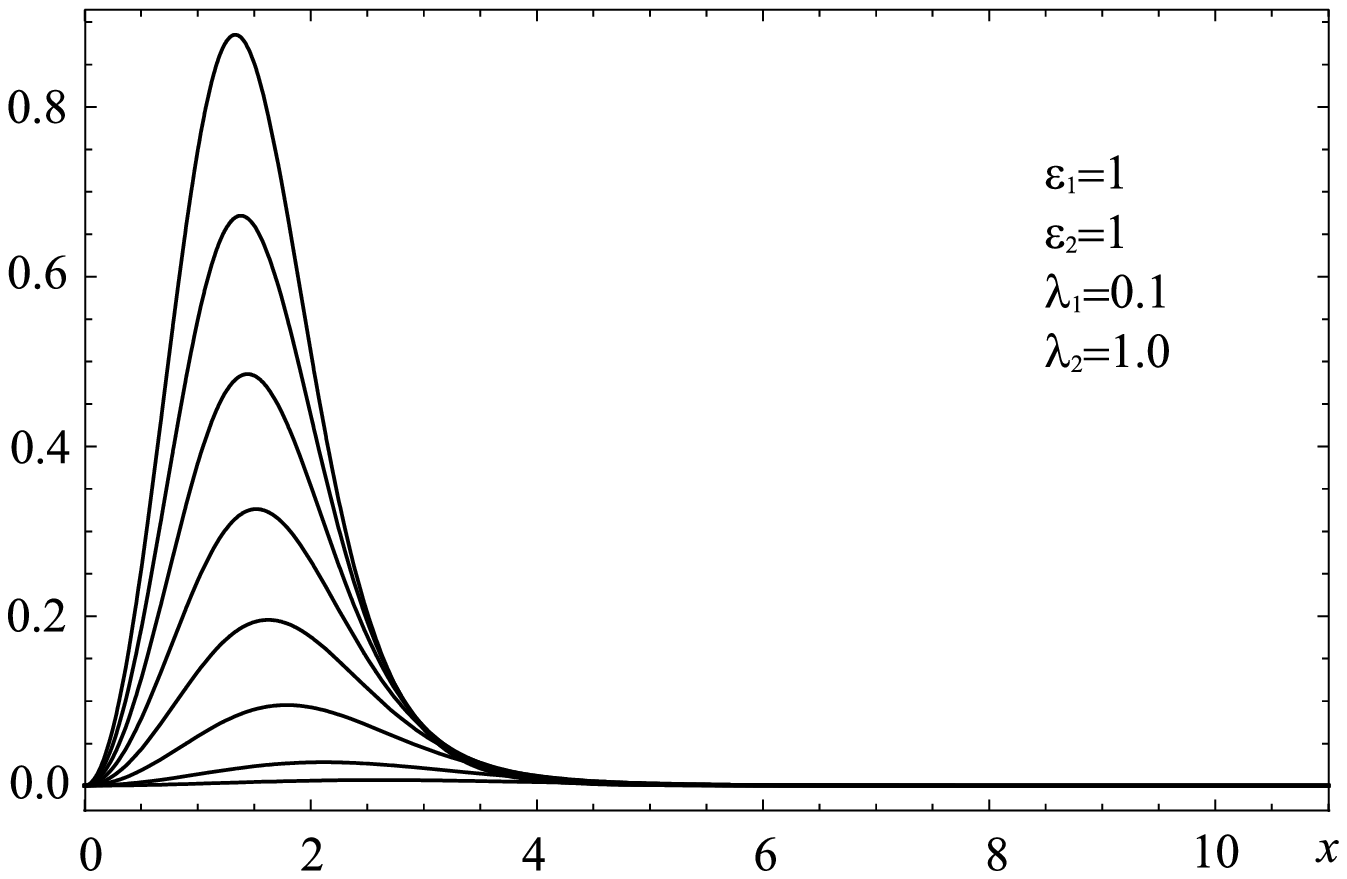}
\vspace{-.2cm}
\caption{\small The dimensionless energy density
 $\varepsilon$ of the system   \eqref{sf1_fix}-\eqref{sf4_fix} from \eqref{energy_dens} for the different initial values of
 $\chi_0$ taken from table \ref{tabl2}. The top line corresponds to the greatest
 $\chi_0=\sqrt{1.4}$, and the bottom one -- to the lowest
 $\chi_0=0.3$.}
  \label{energy_fig}
  \end{center}
\end{minipage}\hfill
\end{figure}

\begin{figure}[t]
\centering
  \includegraphics[height=8cm]{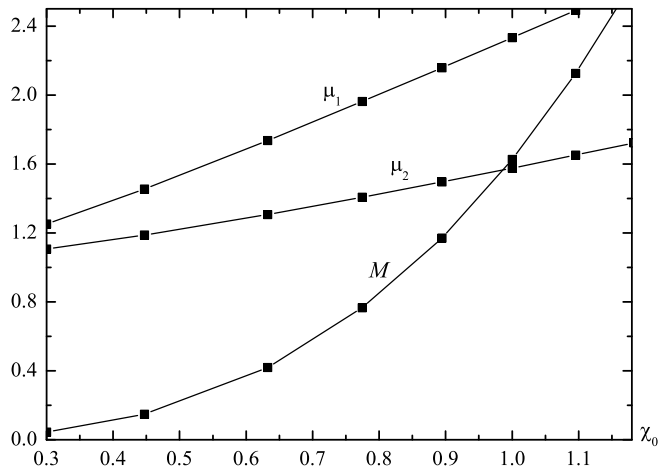}
\vspace{-.8cm}
\caption{\small The dependence of the parameters of the system
 $\mu_1, \mu_2$ and the total energy
 $M$ on the initial values of
 $\chi_0$   for the system \eqref{sf1_fix}-\eqref{sf4_fix}.
The data are taken from  table
 \ref{tabl2}.}
  \label{depend_chi}
\end{figure}

\begin{figure}[t]
\begin{minipage}[t]{.49\linewidth}
  \begin{center}
  \includegraphics[height=5.5cm]{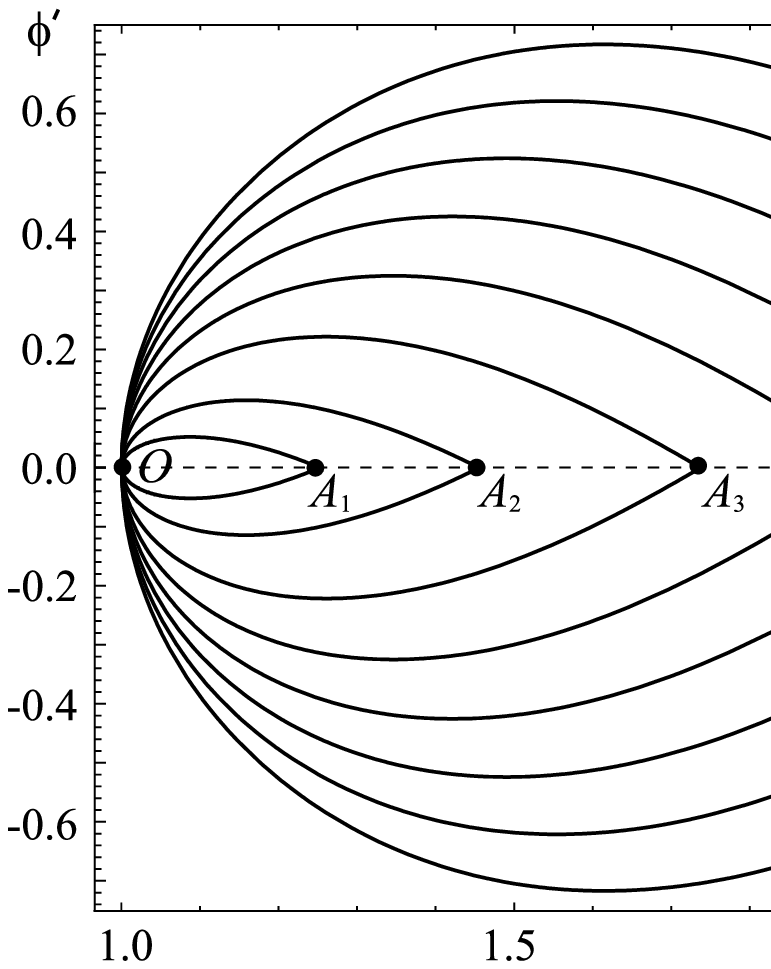}
\vspace{-.2cm}
\caption{\small The phase portrait of the system
 \eqref{sf1_fix}-\eqref{sf4_fix} for the scalar field
 $\phi$ plotted using the parameters from  table
\ref{tabl2}. The points $A_i$ correspond to the fixed point
 $A$ from \eqref{fix1}.
 All solutions go to these points at $x\to \pm \infty$ starting with the different initial values at the point $O$
 corresponding to the origin of coordinates $x=0$. The index
 $i$ runs over $i=1..8$ in accordance with numeration from  table
 \ref{tabl2}.
}
    \label{phasa_phi_fig}
  \end{center}
\end{minipage}\hfill
\begin{minipage}[t]{.49\linewidth}
  \begin{center}
  \includegraphics[height=5.5cm]{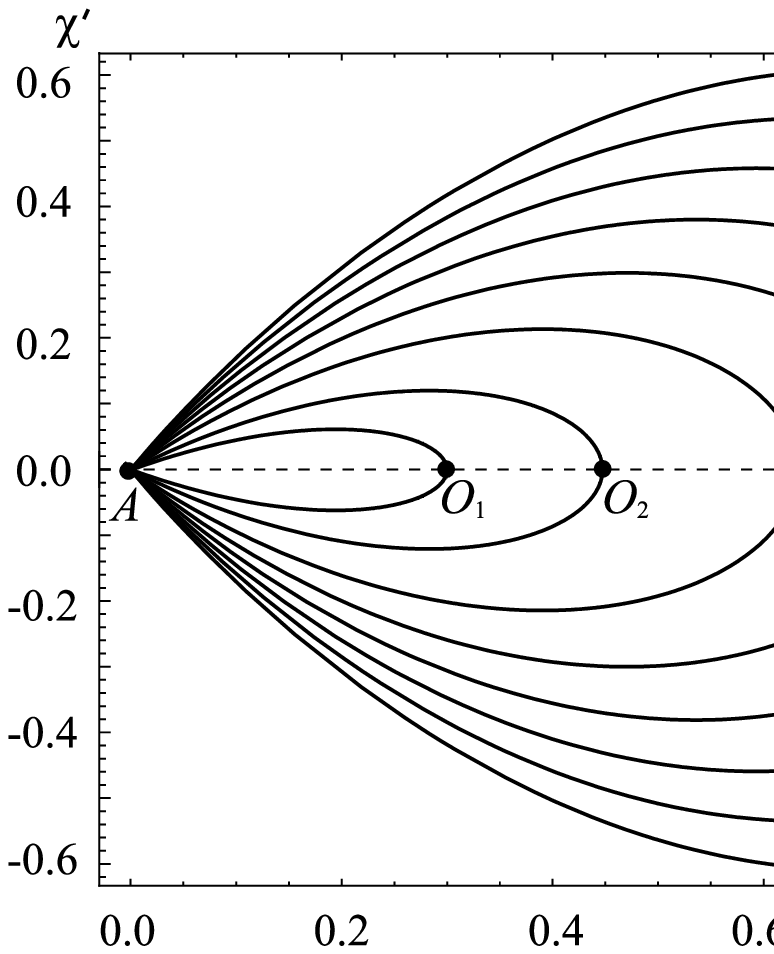}
\vspace{-.2cm}
\caption{\small The phase portrait of the system  \eqref{sf1_fix}-\eqref{sf4_fix} for the scalar field $\chi$
plotted  using the parameters from  table \ref{tabl2}.
The points $O_i$ correspond to the origin of coordinates
 $x=0$ with the different initial values taken from  table \ref{tabl2}.
Asymptotically,  as $x\to \pm \infty$,
the solutions tend to the fixed point
 $A$ from \eqref{fix1} where $\chi(\pm \infty)=0$. The index
 $i$ runs over $i=1..8$ in accordance with numeration from  table \ref{tabl2}.
}
    \label{phasa_chi_fig}
  \end{center}
\end{minipage}\hfill
\end{figure}

All the fixed points, being the stationary points of the system
 \eqref{sf1_fix}-\eqref{sf4_fix}, are situated at $x=\pm \infty$. Then static solutions, if they exist, should start from these points.
One of physically interesting  problems is a $Z_2$ symmetric solution. In this case the symmetry plane is chosen at
$x=0$ where the derivatives
 $\phi^\prime(0)=\chi^\prime(0)=0$. Examples of such solutions are presented in Fig.
 \ref{scal_fields}.  Using the energy-momentum tensor
 \eqref{emt_two_fields} and the above dimensionless variables,
 the dimensionless energy density can be derived   in the following form
\begin{equation}
\label{energy_dens}
\varepsilon=\frac{T_0^0}{\varphi_0^4 \Lambda_3} =
\frac{\epsilon_1}{2}\phi^{\prime 2}+\frac{\epsilon_2}{2}\chi^{\prime 2}+\frac{\lambda_1}{4}\left(\phi^2-\mu_1^2\right)^2
+\frac{\lambda_2}{4}\left(\chi^2-\mu_2^2\right)^2+\frac{1}{2}\phi^2\chi^2-\frac{\lambda_2}{4}\mu_2^4,
\end{equation}
where the constant $\text{const}$ from the potential \eqref{poten_sf} is chosen  equal to
 $-(\lambda_2/4) \mu_2^4$ to make the energy density to be equal to zero at infinity.
Using this expression, one can plot the corresponding graphs for the energy presented in Fig.
 \ref{energy_fig}. The total energy (mass) of the system is defined by the expression
$$
M=\int_{-\infty}^\infty \varepsilon(x)dx.
$$
Calculating this integral, the values of the total energy presented in the last column of table \ref{tabl2} have been found.
Also, using the table, one can plot the graphs of dependence  of the parameters $\mu_1, \mu_2$  and the total energy $M$ on
the initial values of $\chi_0$ presented in Fig. \ref{depend_chi}.
The corresponding phase portraits are shown in Figs.
 \ref{phasa_phi_fig},\ref{phasa_chi_fig}.

\begin{figure}[t]
\centering
  \includegraphics[height=8cm]{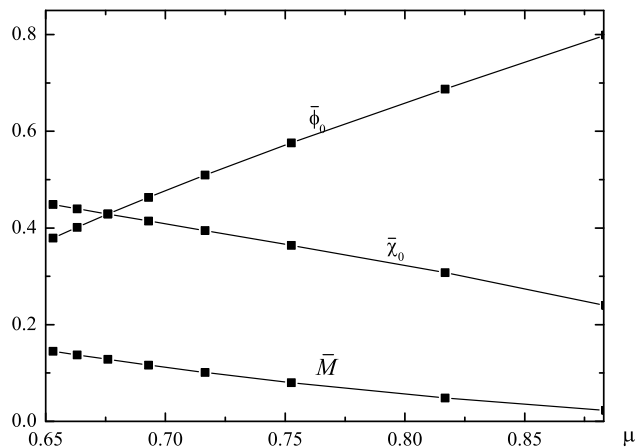}
\vspace{-1cm}
\caption{\small The  dependence of the initial values
 $\bar{\phi}_0, \bar{\chi}_0$ and the total energy
 $\bar{M}$ on the value of the parameter
 $\mu$ for the system \eqref{sf1_fix_n}-\eqref{sf4_fix_n}. The data are taken from  table
 \ref{tabl3}.}
  \label{depend_mu}
\end{figure}

The system \eqref{sf1}-\eqref{sf2} allows introducing another dimensionless variables. Namely, introducing the
dimensionless variables
 $\bar{\phi}=\varphi/m_1, \bar{\chi}=\psi/m_1, \bar{x}=\sqrt{\Lambda_3}m_1 r, \mu=m_2/m_1$,
and also
$\lambda_1=\Lambda_1/\Lambda_3, \lambda_2=\Lambda_2/\Lambda_3$,
one can rewrite this system as follows:
\begin{eqnarray}
\label{sf1_fix_n}
&&\bar{\phi}^\prime=\bar{z},
\\
\label{sf2_fix_n}
&&\bar{\chi}^\prime=\bar{v},
\\
\label{sf3_fix_n}
&&\bar{z}^\prime=\frac{1}{\epsilon_1}\bar{\phi}\left[\bar{\chi}^2+\lambda_1\left(\bar{\phi}^2-1\right)\right],
\\
\label{sf4_fix_n}
&&\bar{v}^\prime=\frac{1}{\epsilon_2}\bar{\chi}\left[\bar{\phi}^2+\lambda_2\left(\bar{\chi}^2-\mu^2\right)\right].
\end{eqnarray}
The values of the parameter
 $\mu$ and initial conditions
 $\bar{\phi}(0)$  and $\bar{\chi}(0)$ at which regular solutions do exist can be obtained from the values presented in  table
\ref{tabl2} by corresponding rescaling the variables: $\bar{\phi}_0=\phi_0/\mu_1, \bar{\chi}_0=\chi_0/\mu_1,
\mu=\mu_2/\mu_1, \bar{x}=\mu_1 x, \bar{M}=M/\mu_1^3$.
The corresponding new values of the mentioned parameters are presented in  table
 \ref{tabl3}. Using this table, one can plot the graphs of dependence of the initial values
 $\bar{\phi}_0, \bar{\chi}_0$ on  $\mu$ presented in Fig.
 \ref{depend_mu}.

\begin{table}[htbp]
 \caption{\small The initial values  $\bar{\phi}_0, \bar{\chi}_0$ and corresponding to them values of the parameter
 $\mu$ for the system
 \eqref{sf1_fix_n}-\eqref{sf4_fix_n}.}
\vspace{-.7cm}
\begin{center}
\begin{tabular}{|p{2.5cm}|p{2.5cm}|p{2.5cm}|p{2.5cm}|}
\multicolumn{4}{c}{} \\[10pt]\hline
$\bar{\phi}_0$ & $\bar{\chi}_0$ & $\mu$ &  $\bar{M}$ \\
\hline
0.799332&0.239799& 0.883765&0.0225663\\
\hline
0.687528&0.307472& 0.816713&0.0481829 \\
\hline
0.575949&0.364262& 0.752563&0.0800622 \\
\hline
0.509456&0.394623& 0.71655&0.101201 \\
\hline
0.463382&0.414461& 0.692895&0.116487 \\
\hline
0.428791&0.428791& 0.675965&0.128174\\
\hline
0.401476&0.439795& 0.66316&0.137485\\
\hline
0.379136&0.4486& 0.653091&0.145127\\
\hline
\end{tabular}
\end{center}
\label{tabl3}
\end{table}

Summarizing, here we have considered
the system with two non-gravitating coupled scalar fields in cartesian coordinates. For such a system, as well as in general relativity,
the task of finding regular solutions  amounts to searching
eigenvalues of the parameters of the model.
The model contains six available parameters: two initial values of the scalar fields $\varphi_0,\psi_0$ and four
free parameters $\Lambda_1, \Lambda_2, m_1, m_2$ ($\Lambda_3$ can be always excluded by redefinition of the parameters
$\Lambda_1, \Lambda_2$, and the initial values of derivatives $\varphi^\prime(0),\psi^\prime(0)$ are chosen to be equal to zero
for obtaining $Z_2$ symmetric solutions). Then for obtaining regular solutions it is necessary to find eigenvalues of only two
of these six parameters.
For example, we have been sought the eigenvalues of the parameters
  $\mu_1$ and $\mu_2$ in the system \eqref{sf1_fix}-\eqref{sf4_fix}. For the obtained eigenvalues regular solutions start at
 $x=0$ and tend to the fixed point  $A$ corresponding to the local minimum of the system (for the case of usual scalar fields
 considered here, see Figs. \ref{scal_fields}, \ref{phasa_phi_fig} and \ref{phasa_chi_fig}), and to the local maximum
 (for phantom/ghost fields). One can see from Fig.
 \ref{depend_chi} that there exist some lowest eigenvalues of the parameters
 $\mu_1$ and $\mu_2$ at which the total energy of the system
 $M$ goes to zero. This corresponds to the existence of some critical
 $\mu_1$ and $\mu_2$ at which physically sensible solutions with a nonzero total energy still exist.
 Similarly, for the system  \eqref{sf1_fix_n}-\eqref{sf4_fix_n} there exists some critical eigenvalue of the parameter
 $\mu$ at which $M\to 0$ as well
 (see Fig. \ref{depend_mu}).

Type ``unstable node'' fixed points allow the existence of sets of solutions starting from these points at
 $x=\pm \infty$ both for usual fields
(the fixed points $A, B$ and $C, D$) and for phantom/ghost scalar fields
 (the point $E$). Thus the qualitative analysis shows that from the point of view of a possibility of obtaining regular localized
 solutions the system with two coupled scalar  fields in question  seems to be quite perspective.
The previous studies from Refs. \cite{Dzhunushaliev:2006sy}-\cite{Dzhunushaliev:2008hq} show that
inclusion of gravitational fields does not change the qualitative behavior of solutions.
Then one would expect, for example, that in the presence of gravitational fields
the lower restriction on the values of the parameters  $\mu_1$ and $\mu_2$ giving physically sensible solutions will also exist.

\end{document}